\documentclass[a4paper, 12pt]{article}
\usepackage[russian]{babel}
\usepackage[cp866]{inputenc}
\usepackage{graphicx}
\usepackage{amsfonts}
\usepackage{amsmath}
\usepackage{amssymb}

\begin{document}

\baselineskip 8.0mm
\thispagestyle{empty}
{\small \it ISSN 1063-7737, Astronomy Letters, 2013, Vol. 39, No. 3, pp. 141-149. 
\copyright Pleiades Publishing, Inc., 2013.
Original Russian Text \copyright N. P. Pitjev, E. V. Pitjeva, 2013, published in 
Pis'ma v Astronomicheskii Zhurnal, 2013, Vol. 39, No. 3, pp. 163-172.}

\bigskip
\bigskip
\centerline{\large\bf Constraints on Dark Matter in the Solar System}
\bigskip
\centerline{\bf N. P. Pitjev$^{1*}$ and E. V. Pitjeva$^{2**}$}
\centerline{\it $^1$St. Petersburg State University, Universitetski pr. 28, St. 
Petersburg, 198504 Russia}
\centerline{\it $^2$Institute of Applied Astronomy of Russian Academy of Sciences, 
nab. Kutuzova 10,} 
\centerline{\it St. Petersburg, 191187 Russia}
\centerline{\small Received July 12, 2012}

\bigskip\noindent

{\bf Abstract} -- We have searched for and estimated the possible gravitational 
influence of dark matter in the Solar system based on the EPM2011 planetary 
ephemerides using about 677 thousand positional observations of planets and 
spacecraft. Most of the observations belong to present-day ranging measurements. 
Our estimates of the dark matter density and mass at various distances from the 
Sun are generally overridden by their errors ($\sigma$). This suggests that the 
density of dark matter ${\rho}_{dm}$, if present, is very low and is much less 
than the currently achieved error of these parameters. We have found that 
${\rho}_{dm}$ is less than $1.1\cdot10^{-20}$ g cm$^{-3}$ at the orbital distance 
of Saturn, ${\rho}_{dm}<1.4\cdot10^{-20}$ g cm$^{-3}$ at the orbital distance of 
Mars, and ${\rho}_{dm}<1.4\cdot10^{-19}$ g cm$^{-3}$ at the orbital distance of the 
Earth. We also have considered the case of a possible concentration of dark 
matter to the Solar system center. The dark matter mass in the sphere within 
Saturn's orbit should be less than $1.7\cdot10^{-10} M_{\odot}$ even if its 
possible concentration is taken into account.

{\bf DOI}: 10.1134/S1063773713020060


\vskip 3em \noindent

\vfill\noindent

$^*$E-mail: {\tt ai@astro.spbu.ru}

$^{**}$E-mail: {\tt evp@ipa.nw.ru}

\newpage
\centerline{INTRODUCTION}
\smallskip

At present, the subject of dark matter attracts rapt attention of physicists and 
astronomers. It is one of the most popular in theoretical and observational works 
concerning cosmology and studies of galactic structures. Dark matter in galaxies 
has long been discussed in stellar dynamics. Its existence was suggested by the 
virial paradox concerning galaxy clusters (Zwicky 1933; Karachentsev 1966) and a 
flat rotation curve for many spiral galaxies. For an explanation, it was 
hypothesized that an additional invisible mass was in the halos of galaxies and 
its value could exceed the visible one by several times. Massive halos (Flynn et 
al. 1996; Karachentsev 2001; Fridman and Khoperskov 2011) are generally included 
when describing galactic structures and in galactic models.

In present-day cosmological theories, it is hypothesized
that the bulk ($\sim 73\%$) of the mean density
of the Universe is accounted for by dark energy,
while about 4\% and 23\% of the remaining part are
accounted for by baryonic and dark matter, respectively
(Kowalski et al. 2008; Komatsu et al. 2011;
Keisler et al. 2011). The dark matter is deemed
to be nonbaryonic in nature and its properties are
speculative. It is believed that this matter is formed
not from atoms, does not interact with ordinary matter
through electromagnetic forces, and its particles
carry no electric charge. Various hypothetical and
exotic particles are proposed as candidates for dark
matter (see the review by Bertone et al. (2005) and
Peter (2012)). If the hypothesis about particles is
correct, then our Galaxy, along with other galaxies,
is immersed in a halo of such dark matter particles,
and these particles with exceptional penetrability
must be everywhere, including the Solar system
and the Earth. Although the particles interact
very weakly with matter, attempts are made to find
them by rare interactions with atoms of ordinary
matter. Experiments using special detectors and telescopes
(CRESST, CoGeNT, DAMA, XENON100,
PAMELA, FERMI, HESS, CDMS, ANTARES,
WMAP, SPT, etc.) are carried out to find and
investigate dark matter particles or traces of their
interaction, and their data are carefully analyzed. New
experiments are planned. The hypothetical particles
can interact with ordinary matter through their
elastic scattering by atomic nuclei (Goodman and
Witten 1985), and various experiments are conducted
to find this effect: CDMS~II, Xenon100, Zeplin~III,
etc. The goal of these studies is an attempt to detect
and measure the number of outlier events per unit
energy, their time and angular dependences. These
quantities are assumed to depend on the local density
and velocity distribution of dark matter particles.

Despite the possible absence or very weak interaction
of dark matter with ordinary one, the dark matter
must have gravitational properties. Since it can be in
the Solar system, the dark matter can manifest itself
through its gravitational influence on Solar system
bodies.

Attempts to detect the influence of possible dark
matter on the motion of bodies in the Solar system
have already been made. Nordtwedt (1994) and
Nordtwedt et al. (1995) used laser observations of
the Moon and found an upper limit for the possible
acceleration in the presence of dark matter: $3\cdot10^{-14}$ 
cm s$^{-2}$ There are several works where
the effects are searched for in the motions of planets
and other bodies in the Solar system (Anderson
et al. 1989, 1995; Khriplovich and Pitjeva 2006; Iorio
2006: Khriplovich 2007; Fre` re et al. 2008). Table 1
lists previous estimates of the dark matter density ${\rho}_{dm}$
and ${M_{dm}}$ in the Solar system. The third column
gives the distance $r$ in astronomical units (AU) from
the Sun corresponding either to the distance at which
the density ${\rho}_{dm}$ was estimated or the radius of the
sphere within which the mass ${M_{dm}}$ was estimated.
The goal of this paper is an attempt to detect the
gravitational manifestation of dark matter or to give
a constraining upper limit for the dark matter density
and mass in the Solar system using a new version
of the planetary ephemerides, EPM2011, and new
observations of planets and spacecraft.

\smallskip
{\bf Table 1.} Estimates of the dark matter density and mass in the Solar system

 %\bigskip
\begin{tabular}{|c|c|c|c|c|}
\noalign{\smallskip}
\hline
\noalign{\smallskip}
Year & Authors & Distance $r$, & Density ${\rho}_{dm}$, & Mass \\
   &     &  AU &  g cm$^{-3}$ & $M(r)_{dm}$ ($M_{\odot}$) \\
\noalign{\smallskip}
\hline
\noalign{\smallskip}
1989 & Anderson et al. & 19.2 & & $<3\cdot10^{-6}$ \\[-2pt]
\hline
\noalign{\smallskip}
1995 & Anderson et al. & 19.2 & & $<5\cdot10^{-7}$  \\[-2pt]
     &                 & 30.1 & & $<3\cdot10^{-6}$  \\[-2pt]
\noalign{\smallskip}
\hline
\noalign{\smallskip}
1996  & Gron and Soleng & 1.08 & $<1.8\cdot10^{-16}$  &   \\ [-2pt]
      &             & 19.2 &  & $<2\cdot10^{-6}$ \\ [-2pt]
\noalign{\smallskip}
\hline
\noalign{\smallskip}
2006  & Khriplovich and Pitjeva & 1.52 & $<3\cdot10^{-19}$ &  \\[-2pt]
\noalign{\smallskip}
\hline
\noalign{\smallskip}
2006 & Iorio & 1.52  & $<4\cdot10^{-19}$  &    \\[-2pt]
\noalign{\smallskip}
\hline
\noalign{\smallskip}
2006 & Sereno and Jetzer & 1.52 & $<3\cdot10^{-19}$ &   \\[-2pt]
\noalign{\smallskip}
\hline
\noalign{\smallskip}
2008 & Fr\`{e}re et al. & 1.52 & $<3\cdot10^{-19}$ &  \\[-2pt]
\noalign{\smallskip}
\hline
\noalign{\smallskip}
\end{tabular}
 
\smallskip

\bigskip
\centerline{POSSIBLE OBSERVATIONAL EFFECTS}
\smallskip

If dark matter is present in the Solar system, then
it should lead to some additional gravitational influence
on all bodies. The effect will depend on the
density of dark matter, on its distribution in space,
etc. Let us assume, as is usually done (Anderson
et al. 1989, 1995; Gron and Soleng 1996; Khriplovich
and Pitjeva 2006; Sereno and Jetzer 2006; Fr\`{e}re
et al. 2008) that dark matter of an unknown nature
is distributed in the Solar system spherically symmetrically
relative to the Sun. Apart from the already
accountable accelerations from the Sun, planets,
asteroids, and trans-Neptunian objects (TNOs),
any planet at distance $r$ from the Sun can then be
assumed to undergo an additional acceleration from
dark matter:

\begin{equation}\label{f-1}
   {\bf\ddot r}_{dm} = - {GM(r)_{dm}\over{r^3}} \bf r  ,
\end{equation}
where $M(r)_{dm}$ is the mass of the additional matter in
a sphere of radius $r$ around the Sun. 

Thus, if we assume that there is an extended gravitating
medium in the Solar system, then when finding
the central attractive mass (or the correction to the
heliocentric gravitational constant $GM_{\odot}$) from observational
data separately for each planet, we would
obtain an increasing value of this mass in accordance
with the additional mass within the sphere with the
mean radius of the planetary orbit. With sufficiently
accurate observational data and a sufficiently large
amount of interplanetary matter, this dependence of
the central attractive mass on the semimajor axis of
the planetary orbit not only could be an indicator for
the presence of dark matter but also could characterize
the increasing amount of additional mass with
distance from the Sun, i.e., the density distribution.
In particular, the processing high-accuracy observations
for Mars and Saturn located at different distances
from the Sun (1.52 and 9.58 AU, respectively)
could provide data on the presence or absence of an
appreciable amount of dark matter between the orbits
of these planets.

Another consequence of the presence of a continuous
gravitating medium in interplanetary space is its
influence on the motion of the perihelion of a planetary
orbit. At a uniform density ${\rho}_{dm}$ of the gravitating
medium filling the Solar system, the additional acceleration
on a body will be proportional to $r$:

\begin{equation}\label{f-2}
   {\bf\ddot r}_{dm} = - k\bf r  ,
\end{equation}
where $k$ is a constant coefficient related to the density
of the medium ${\rho}_{dm}$: $k=4/3{\pi}G{\rho}_{dm}$, $G$ is the gravitational
constant. In other cases, the dependence on $r$
is more complex. If we denote the energy and area
integrals per unit mass by $E$ and $J$ and a spherically
symmetric potential by $U(r)$, then (Landau and Lifshitz
1988) the equation of motion along the radius $r$
can be written as

\begin{equation}\label{f-3}
   \dot r = ( 2[E+U(r)] - (J/r)^2 )^{1/2} ,
\end{equation}
and the equation along the azimuth $\theta$ is

\begin{equation}\label{f-4}
    {d \theta \over dr} = {{J/r^2} \over ( 2[E+U(r)] - (J/r)^2 )^{1/2}} .
\end{equation}

In the Newtonian two-body problem, the oscillation
periods along the radius $r$ (from the pericenter
to the apocenter and back) and azimuth $\theta$ around
the center coincide, and the positions of the pericenters
and apocenters are not shifted from revolution
to revolution. In the general case of a spherically
symmetric potential different from the central field of
a point mass or a homogeneous sphere, the bounded
trajectory is not closed and fills everywhere densely
the flat ring between the pericenter and apocenter
distances. Since the trajectory is not closed, the
pericenter and apocenter positions are shifted from
revolution to revolution:

\begin{equation}\label{f-5}
{\theta}_1-{\theta}_0 = 2\int_a^b {{J/r^2} \over (2[E+U(r)]-(J/r)^2 )^{1/2}}dr .
\end{equation}
where $a, b$ are the minimum and maximum radial distances,
${\theta}_0, {\theta}_1$ are the initial and next positions of the
pericenter. The presence of an additional gravitating
medium leads to a shorter radial period and a negative
drift of the pericenter and apocenter positions (in a
direction opposite to the planetary motion). The perihelion
drift for uniformly distributed matter (${\rho}_{dm}$ =
const) depends on the orbital semimajor axis $a$ and
eccentricity $e$ (Khriplovich and Pitjeva 2006):

\begin{equation}\label{f-5a}
\Delta {\theta}_0 = -4{\pi}^2{\rho}_{dm}a^3(1-e^2)^{1/2}/M_{\odot},
\end{equation}
where $\Delta {\theta}_0$ is the perihelion drift in one complete radial
oscillation. Since the eccentricity $e$ for the planets in
the Solar system is small in most cases, the dependence
on $e$ is occasionally neglected in Eq. (\ref{f-5a}) for the
perihelion drift, as was done in Fr\`{e}re et al. (2008).

It should be taken into account that the Solar system
has its own extended medium associated with the
solar wind and plasma. The solar wind produces an
almost spherically symmetric distribution of the particle
flux (Parker 1963) whose space density decreases
rapidly with increasing distance from the Sun, becoming
vanishingly small on the periphery. Data from
interplanetary spacecraft revealed that the solar wind
particle flux density changes approximately as $r^{-2}$,
where $r$ is the distance from the Sun, up to the orbit
of Jupiter (Parker 1963, 1968; Hundhausen 1972).
The solar wind density is $10^{-23}$ g cm$^{-3}$ near the
Earth's orbit and decreases to $10^{-25}$ gcm$^{-3}$ at
the distance of Saturn. The total mass of the solar
wind plasma up to Saturn's orbit is approximately
$10^{-15}$. These values are too small to be detected
at present. Provided that the dark matter exceeds
appreciably these estimates, it becomes possible to
find its manifestations and to separate its effects from
the medium with ordinary properties associated with
the Solar system.

The density and mass of the dark matter are more
commonly estimated by assuming that it changes
very slowly or is constant within the Solar system,
i.e., by assuming its distribution to be uniform. The
concentration of dark matter to the center and even
its capture and direct fall to the Sun are assumed in
a number of papers (Lundberg and Edsj\"{o} 2004; Peter
2009; Iorio 2010). The latter assumptions should
be made with caution. Pitjeva and Pitjev (2012)
found a secular decrease in the heliocentric gravitational
constant $GM_{\odot}$: $\dot{GM}_{\odot}/GM_{\odot} \simeq (-5.0)\cdot10^{-14}$
per year. This is primarily due to the decrease
in solar mass through radiation and the solar
wind. Therefore, there is a stringent constraint on the
amount of possible dark matter falling to the Sun. In
any case, it is less than that assumed by Iorio (2010)
by several orders of magnitude. A serious constraint
on the possible presence of dark matter inside the
Sun (no more than 2-5\% of the solar mass) was also
obtained by Kardashev et al. (2005), who carefully
analyzed the physical characteristics of the Sun.

Whereas the integral effect of the entire additional
mass in the volume up to the planetary orbit is important
in estimating the change of the central attractive
mass, the local effect of the gravitational field
difference near the planetary orbit due to the presence
of an additional gravitating medium is important in
searching for an additional change in the perihelion
position. The additional planetary perihelion precession
is investigated by taking into account all other
known effects affecting the perihelion drift. Note that
in the case of a small change in the central mass
with time, there is no precession of the pericenter
and apocenter positions (Pitjeva and Pitjev 2012),
but if there is also an additional gravitating medium,
then a negative drift of the perihelion and aphelion
occurs from revolution to revolution in accordance
with Eqs. (\ref{f-5}) and (\ref{f-5a}). Since the growth of the perihelion
shift is accumulated, this criterion (effect) can
be fairly sensitive for testing the presence of additional
matter.

\bigskip
\centerline{OBSERVATIONAL DATA AND THEIR PROCESSING}
\smallskip

Finding the effects related to the possible presence
of dark matter in the Solar system requires using
highly accurate observations and a careful allowance
for other small effects that may turn out to be comparable
to the sought-for ones. For example, the
weak effect from solar oblateness on the motion of
Mercury and on the drift of its perihelion may turn
out to be of the same order of magnitude with the
action of dark matter. Different parameters of the
planetary ephemerides are estimated from processing
observations of different types, from classical
meridian ones to present-day radio and spacecraft
observations (Pitjeva 2008). Here, we use the optical
observations since 1913, when an improved micrometer
was installed at the US Naval Observatory
and the observations became more accurate (about
$0\rlap.''5$), up to all the available present-day observations
in 2011. It should be noted that the accuracies of
present-day optical CCD observations and spacecraft
trajectory observations reach, respectively, a
few hundredths of an arcsecond and a few meters
(at the distance of Saturn). Most of the observations
were retrieved from the database of the US
Jet Propulsion Laboratory (JPL) created by E.M.
Standish and updated and maintained at present by
W.M. Folkner: http://iau-comm4.jpl.nasa.gov/plan-eph-
data/index.html. These data were supplemented
with the Russian radar observations of planets
(1961-1995), http://www.ipa.nw.ru/PAGE/ DEPFUND/
LEA/ENG/rrr.html, and with the Venus Express
and Mars Express data obtained due courtesy
of A. Fienga. The volume of highly accurate observations
on which the next EPM (Ephemerides of
Planets and the Moon) versions are based increases
continuously, and the total number of observations
used in the current version of the EPM2011 planetary
theory is 676 804 (Table 2).

\smallskip
{\bf Table 2.} Observational material

\begin{tabular}{|l|c|c|c|c|}
\noalign{\smallskip}
\hline
\noalign{\smallskip}
Planet &\multicolumn{2}{|c|}{Radio} &\multicolumn{2}{|c|}{Optical}\\
\cline{2-5}
 & interval & number & interval & number \\
\noalign{\smallskip}
\hline
\noalign{\smallskip}
Mercury & 1964-2009 & 948 & -- & -- \\[-2pt]
\noalign{\smallskip}
\hline
\noalign{\smallskip}
Venus  & 1961-2010 & 40281 & -- & -- \\ [-2pt]
\noalign{\smallskip}
\hline
\noalign{\smallskip}
Mars   & 1965-2010 & 578918 & -- & -- \\[-2pt]
\noalign{\smallskip}
\hline
\noalign{\smallskip}
Jupiter + 4 satellites & 1973-1997 & 51 & 1914-2010 & 13023 \\[-2pt]
\noalign{\smallskip}
\hline
\noalign{\smallskip}
Saturn + 9 satellites & 1979-2009 & 126 & 1913-2010 & 14744 \\[-2pt]
\noalign{\smallskip}
\hline
\noalign{\smallskip}
Uranus + 4 satellites & 1986 & 3 & 1914-2010 & 11681 \\[-2pt]
\noalign{\smallskip}
\hline
\noalign{\smallskip}
Neptune + 1 satellite & 1989 & 3 & 1913-2010 & 11474 \\[-2pt]
\noalign{\smallskip}
\hline
\noalign{\smallskip}
Pluto & -- & -- & 1914-2010 & 5552 \\[-2pt]
\noalign{\smallskip}
\hline
\noalign{\smallskip}
{\bf Total} &  &{\bf 620330} & & {\bf 56474} \\
\noalign{\smallskip}
\hline
\noalign{\smallskip}
\end{tabular}
 
\smallskip

The observations were processed using proved and
tested techniques by taking into account all of the
necessary reductions (Pitjeva 2005).

The reductions of the radar observations:
\begin{itemize}
\itemsep -2mm
\item the reduction of the instants of time to a uniform scale;
\item the relativistic corrections - the time delay in
the gravitational fields of the Sun, Jupiter, and
Saturn (Shapiro effect) and the transition from
the coordinate time (ephemeris argument) to
the observer's proper time;
\item the time delay in the Earth's troposphere;
\item the time delay in the solar coronal plasma;
\item the correction for the planetary surface topography
(Mercury, Venus, Mars).
\end{itemize}

 The reductions of the optical observations:
\begin{itemize}
\itemsep -2mm
\item the reduction of the observations to the ICRF:
the reference catalogs => FK4 => FK5  => ICRF;
\item the correction for the additional phase effect;
\item the correction for the gravitational deflection of
light by the Sun.
\end{itemize}

Present-day radio observations of planets and
spacecraft with a 1-m accuracy (a relative error
of $10^{-12} \div 10^{-11}$)) make it possible to estimate very
subtle and small (in magnitude) effects in the Solar
system (see, e.g., Konopliv et al. 2011; Pitjeva 2010;
Fienga et al. 2011). Substantial progress is related
to several factors: an increase in the accuracy of observational
data reduction procedures and dynamical
models of motion and an improvement in the quality
of observational data, an increase in their accuracy
and the extent of the time interval on which these
observations were obtained.

\bigskip
\centerline{THE EPM2011 PLANETARY EPHEMERIDES}
\smallskip

The EPM2011 (Ephemerides of Planets and the
Moon) numerical ephemerides were constructed using
about 677 thousand observations (1913-2010)
of various types. The equations of motion for bodies
were taken for a parameterized post-Newtonian n-body
metric. The integration in the barycentric coordinate
system in the TDB scale at epoch J2000.0
was performed by Everhart's method in an interval
of 400 years (1800-2200) by the lunar-planetary
integrator of the ERA-7 software package (Krasinsky
and Vasilyev 1997). The EPM ephemerides,
along with the corresponding TT-TDB time differences
and the coordinates of seven additional objects
(Ceres, Pallas, Vesta, Eris, Haumea, Makemake,
Sedna), are accessible via FTP: ftp://quasar.ipa.nw.
ru/incoming/EPM/.

Apart from the mutual perturbations of the major
planets and the Moon, the EPM2011 dynamical
model includes:
\begin{itemize}
\itemsep -2mm
\item the perturbations from 301 most massive asteroids;
\item the perturbations from the remaining minor
planets of the main asteroid belt modeled by a
homogeneous ring;
\item the perturbations from 21 largest trans-Neptunian objects (TNOs); 
\item the perturbations from the remaining trans-Neptunian planets modeled by a 
homogeneous ring at a mean distance of 43 AU;
\item the perturbations from solar oblateness $(2\cdot10^{-7})$;
\item the perturbations caused by the nonsphericity
of the Earth's and Moon's figures;
\item the relativistic perturbations from the Sun, the
Moon, planets and five largest asteroids. 
\end{itemize}

Since the radio measurements where the distances
are predominantly measured were the main observational
material when creating the next version of planetary
ephemerides, EPM2011, controlling the orientation
of the coordinate system for the ephemerides
with respect to the ICRF requires particular attention
and carefulness. The orientation was performed using
VLBI observations of spacecraft near planets against
the background of quasars whose coordinates are
given in the ICRF (Table~3). An example of VLBI
observations for Cassini near Saturn are given in Jones
et al. (2011).

\smallskip
{\bf Table~3.} VLBI observations of spacecraft near planets
against the background of quasars

\begin{tabular}{|c|c|c|c|}
\noalign{\smallskip}
\hline
Spacecraft & Planet & Interval & Number of measurements \\
\noalign{\smallskip}
\hline
\noalign{\smallskip}
Magellan & ‚Ґ­Ґа  & 1990-1994 & 18($\alpha+\delta$) \\[-2pt]
\noalign{\smallskip}
\hline
\noalign{\smallskip}
Venus Express & ‚Ґ­Ґа  & 2007-2010 & 29($\alpha+\delta$) \\[-2pt]
\noalign{\smallskip}
\hline
\noalign{\smallskip}
Phobos & Њ аб & 1989 & 2($\alpha+\delta$) \\[-2pt]
\noalign{\smallskip}
\hline
\noalign{\smallskip}
MGC & Њ аб & 2001-2003 & 15($\alpha+\delta$) \\[-2pt]
\noalign{\smallskip}
\hline
\noalign{\smallskip}
Odyssey & Њ аб & 2002-2010 & 86($\alpha+\delta$) \\[-2pt]
\noalign{\smallskip}
\hline
\noalign{\smallskip}
MRO & Њ аб & 2006-2010 & 41($\alpha+\delta$) \\[-2pt]
\noalign{\smallskip}
\hline
\noalign{\smallskip}
Cassini & ‘ вга­ & 2004-2009 & 22($\alpha,\delta$) \\[-2pt]
\noalign{\smallskip}
\hline
\multicolumn{4}{l}{{\it Note:} \small 
($\alpha+\delta$) denotes one-dimensional measurements of the
combination of $\alpha$ and $\delta$,}\\
\multicolumn{4}{l}{\small ($\alpha, \delta$) denotes two-dimensional measurements.}
%\noalign{\smallskip}
\end{tabular}

\smallskip

The accuracy of such observations improved to
a few tenths of mas (1 mas = $0\rlap.''001$) for Mars and
Saturn in 2001-2010, which allowed the orientation
of the coordinate system for the EPM ephemerides to
be refined (Table~4).

\smallskip
{\bf Table~4.} Rotation angles of the coordinate system for the EPM ephemerides 
in the ICRF (1 mas = $0\rlap.''001$)

\begin{tabular}{|c|c|c|c|c|}
\noalign{\smallskip}
\hline
\noalign{\smallskip}
Interval of observ. & Number of observ. & $\varepsilon_x$, mas & 
$\varepsilon_y$, mas & $\varepsilon_z$, mas \\
\noalign{\smallskip}
\hline
\noalign{\smallskip}
1989-1994 & 20 & $4.5 \pm 0.8$ & $-0.8 \pm 0.6$ & $-0.6 \pm 0.4$ \\[-2pt]
\hline
\noalign{\smallskip}
1989-2003 & 62 & $1.9 \pm 0.1$ & $-0.5 \pm 0.2$ & $-1.5 \pm 0.1$ \\[-2pt]
\noalign{\smallskip}
\hline
\noalign{\smallskip}
1989-2007 & 118 & $-1.528\pm0.062$ & $1.025\pm0.060$ & $1.271\pm0.046$ \\[-2pt]
\noalign{\smallskip}
\hline
\noalign{\smallskip}
1989-2010 & 213 & $-0.000 \pm 0.042$ & $-0.025 \pm 0.048$ & $0.004 \pm 0.028$ \\[-2pt]
\hline
\noalign{\smallskip}
\end{tabular}

\smallskip

More than 260 parameters were determined and
refined in the main version of the improvement of the
planetary part of the EPM2011 ephemerides:

\begin{itemize}
\itemsep -2mm
\item the orbital elements of the planets and satellites
of the outer planets;
\item the astronomical unit or $GM_{\odot}$;
\item the orientation angles of the ephemerides relative
to the ICRF; 
\item the rotation parameters of Mars and the coordinates
of three landers on Mars; 
\item the masses of 21 asteroids, the mean densities
of the taxonomic classes of asteroids (C, S, M); 
\item the mass and radius of the asteroid ring, the
mass of the TNO ring; 
\item the Earth-to-Moon mass ratio; 
\item the quadrupole moment of the Sun and solar
corona parameters for different conjunctions of
the planets with the Sun; 
\item the coefficients of Mercury's topography and
the corrections to the level surfaces of Venus
and Mars; 
\item the coefficients for the additional phase effects
of the outer planets. 
\end{itemize}

The parameters determined while fitting the DE
and EPM ephemerides (Pitjeva and Standish 2009)
and approved by the 27th General Assembly of
the International Astronomical Union in 2009 as
the current best values for ephemeris astronomy
(Luzum et al. 2011) were used in EPM2011 as
the initial ones; they were subsequently improved
based on all observations. Among them, there are
the masses of the largest asteroids:
$M_{Ceres} / M_{\odot}$  = $4.726(8)\cdot10^{-10}$,
$M_{Pallas} / M_{\odot}$  = $1.048(9)\cdot10^{-10}$,
$M_{Vesta} / M_{\odot}$  = $1.297(5)\cdot10^{-10}$;
the Earth-to-Moon mass ratio: $M_{Earth}/M_{Moon} = 81.3005676 \pm 0.00000006 $;
and the astronomical unit in meters: AU $= (149597870695.88 \pm 0.14)$ m.

Special efforts were made for a more accurate
allowance for the overall influence of asteroids on the
motion of planets, most of which are located in the
main asteroid belt. In EPM the main asteroid ring
is modeled by the motion of 301 large asteroids and
a homogeneous material ring representing the influence
of the remaining numerous small asteroids. The
parameters characterizing the asteroid ring of small
asteroids were determined from processing observations:
$$ M_{ring}=(1.06\pm1.12)\cdot10^{-10} M_{\odot},\quad 
R_{ring}=(3.57\pm0.26)\, \hbox{ Ґ}. $$ 
The total mass of the main-belt asteroids represented
by the sum of the masses of 301 largest asteroids and
the asteroid ring is
$ M_{belt} = (12.3 \pm 2.1)\cdot10^{-10} M_{\odot}$ ($\approx 3$ masses of Ceres).
The gravitational perturbations
from TNOs is similarly modeled by 21 known
TNOs and an additional homogeneous ring with a
radius of 43 AU representing the influence of the
remaining smaller objects. The mass of the TNO ring
found from processing observations is
 $$ M_{TNOring} = (501 \pm 249)\cdot10^{-10} M_{\odot} $$
The total mass of all TNOs, including the mass
of Pluto, 21 largest TNOs, and the TNO ring, is
$M_{TNO} = 790\cdot10^{-10} M_{\odot}$, ($\approx 164$ masses of Ceres or
two masses of the Moon).

The uncertainties of parameters of this section
correspond to a $3\sigma$ formal standard error of the least-squares method.

\bigskip
\centerline{RESULTS}
\smallskip

We used the following approach to find and test the
possible effects in the motions of planets.

At the first step, we improved the heliocentric
gravitational constant $GM_{\odot}$ from processing the observations
of all planets. If there is actually an additional
gravitating medium, then the value obtained
will be some mean value with allowance made for the
extended matter. At the next step, having fixed most
of the derived parameters, we processed the observations
for one of the chosen planets and searched
for $GM_{\odot}$ and an additional perihelion shift based on
the observational data only for this planet. Since the
expected dark matter density is low (Table~1), it is
desirable to use the data for more distant planets to
include a larger volume of the influencing invisible
medium in our analysis. Since the expected pattern
of change in the central attractive mass and the perihelion
drift with increasing distance in the presence
of an additional gravitating medium are known, comparison
with the results obtained can be made. If
the values found turned out to be larger than their
errors, then the correspondence of the derived pattern
of change in the central attractive mass and the sign
of the perihelion drift to the expected ones should be
tested. If the errors exceed the values themselves,
then only the upper limit for the possible additional
mass and the density of the distributed matter can be
judged.
 
Tables~5 and 6 list the corrections to the perihelion
precessions and the central mass found from the
observations of planets and spacecraft. 

\smallskip
{\bf Table~5.} Additional perihelion precessions from the observations
of planets and spacecraft

\begin{tabular}{|c|c|c|}
\noalign{\smallskip}
\hline
\noalign{\smallskip}
Planets & $\dot \pi$& |${\sigma_{\dot \pi} / \dot \pi}$| \\
        &  mas/yr &   \\
\noalign{\smallskip}
\hline
\noalign{\smallskip}
Mercury & $-0.020 \pm 0.030$& 1.5 \\[-2pt]
\noalign{\smallskip}
\hline
\noalign{\smallskip}
Venus  & $0.026 \pm 0.016$& 0.62 \\ [-2pt]
\noalign{\smallskip}
\hline
\noalign{\smallskip}
Earth  & $0.0019 \pm 0.0019$& 1.0 \\ [-2pt]
\noalign{\smallskip}
\hline
\noalign{\smallskip}
Mars   & $-0.00020 \pm 0.00037$& 1.9 \\[-2pt]
\noalign{\smallskip}
\hline
\noalign{\smallskip}
Jupiter & $0.587 \pm 0.283$& 0.48  \\[-2pt]
\noalign{\smallskip}
\hline
\noalign{\smallskip}
Saturn & $-0.0032 \pm 0.0047$& 1.5  \\[-2pt]
\noalign{\smallskip}
\hline
\noalign{\smallskip}
\end{tabular}

\smallskip
To control
the derived quantities and their errors, to check the
stability of their values, and to reduce the influence
of possible correlations, we considered various cases
with different numbers of simultaneously determined
parameters. This makes it possible to obtain more
reliable errors ($\sigma$) for the corrections found that better
reflect and correspond to the actual accuracy of the
results obtained. Therefore, the errors in Tables~5
and 6 generally exceed the formal ones by several
times.

\smallskip
{\bf Table~6.} Corrections to the central attractive mass

\begin{tabular}{|c|c|c|}
\noalign{\smallskip}
\hline
\noalign{\smallskip}
Planets & $\Delta M_{0}$& |${\sigma_{\Delta M_{0}} / \Delta M_{0}}$| \\
        & $10^{-10} M_{\odot}$ &\\
\noalign{\smallskip}
\hline
\noalign{\smallskip}
Mercury & $-0.5 \pm 117.7$& 235.4 \\[-2pt]
\noalign{\smallskip}
\hline
\noalign{\smallskip}
Venus  & $-0.67 \pm 5.86$& 8.7 \\ [-2pt]
\noalign{\smallskip}
\hline
\noalign{\smallskip}
Mars  & $0.20 \pm 2.65$& 13.3 \\ [-2pt]
\noalign{\smallskip}
\hline
\noalign{\smallskip}
Jupiter   & $0.4 \pm 1671.4$& 4178.5 \\[-2pt]
\noalign{\smallskip}
\hline
\noalign{\smallskip}
Saturn & $-0.27 \pm 15.16$& 56.1  \\[-2pt]
\noalign{\smallskip}
\hline
\noalign{\smallskip}
\end{tabular}

\smallskip

The values found in both Tables~5 and 6 are generally
overridden by their errors ($\sigma$), indicating that
the density of the dark matter ${\rho}_{dm}$, if present, is very
low and is much less than the currently achieved
error in these parameters. The derived opposite signs
and the absence of general trends in the change of
the corrections themselves to the attractive central
mass and the perihelion precession depending on the
distance from planet to planet also suggest that the
sought-for effects are small.

The relative error in the correction to the central
mass from the observational data separately for
each planet (Table~6) turned out to be considerably
larger than that for the additional perihelion precession
(Table~5) and exceeds the corrections to the
central mass themselves by several times or even
by orders of magnitude. It should be kept in mind
that the accuracy of allowance for and knowledge of
all masses of the bodies that fell into a spherically
symmetric volume relative to the Sun plays a major
role in the integral estimate of the dark matter mass
in this volume. The total amount of dust, meteoric
matter, and solar wind plasma is comparatively small
(less than $10^{-15} \div 10^{-13} M_{\odot}$ in the volume of Saturn's
orbit). Incomplete and inaccurate knowledge of the
asteroid masses play a major role in the uncertainty;
in particular, the error in the mass of the main asteroid
belt is $2 \cdot 10^{-10} M_{\odot}$. The problem of improving the
asteroid masses, their number, and distribution in
the main asteroid belt is topical and important for
increasing the accuracy of planetary theories. More
accurate results were obtained for the perihelion precession
estimates, which allow the local dark matter
density at the mean orbital distance of a planet to
be estimated. Here, the error is comparable to the
values themselves (Table~5) and, therefore, the data
from Table~5 were used for constraining estimates.

To a first approximation, a uniform distribution
can be assumed for the distributed medium, as is
done most often for such estimates, and its density
is then determined from the planets with the most
accurately estimated perihelion precessions that are
farthest from the Sun. Although there is a negative
secular perihelion drift for some of the planets, the
error for all planets is comparable to or appreciably
larger than the absolute values of the derived perihelion
precessions (Table~5). Therefore, attention
should be focused on the errors themselves. The
latter may be considered as an upper limit for the
possible perihelion precession in absolute value |$\delta \pi$| (arcsec yr$^{-1}$)
and, thus, using Eq. (\ref{f-5a}) it can give an
upper limit for the density of the distributed matter.
Our ${\rho}_{dm}$ estimates are given in Table~7.

\smallskip
{\bf Table~7.} Estimates of the density ${\rho}_{dm}$ from the perihelion
precessions

 %\bigskip
\begin{tabular}{|c|c|c|}
\noalign{\smallskip}
\hline
\noalign{\smallskip}
Planets & $\sigma_{\dot \pi}$, arcsec yr$^{-1}$ & ${\rho}_{dm}$, g cm$^{-3}$ \\
\noalign{\smallskip}
\hline
\noalign{\smallskip}
Mercury & $0.000030$ & $<9.3\cdot 10^{-18}$\\[-2pt]
\noalign{\smallskip}
\hline
\noalign{\smallskip}
Venus & $0.000016$ & $<1.9\cdot 10^{-18}$\\[-2pt]
\noalign{\smallskip}
\hline
\noalign{\smallskip}
Earth  & $0.0000019$ & $<1.4\cdot 10^{-19}$\\ [-2pt]
\noalign{\smallskip}
\hline
\noalign{\smallskip}
Mars   & $0.00000037$ & $<1.4\cdot 10^{-20}$\\[-2pt]
\noalign{\smallskip}
\hline
\noalign{\smallskip}
Jupiter & $0.000283$ & $<1.7\cdot 10^{-18}$\\[-2pt]
\noalign{\smallskip}
\hline
\noalign{\smallskip}
Saturn & $0.0000047$ & $<1.1\cdot 10^{-20}$\\[-2pt]
\noalign{\smallskip}
\hline

\end{tabular}

\smallskip

The estimates from the data for the Earth, Mars,
and Saturn give the most stringent constraints on
the density. If we proceed from the assumption of a
uniform ${\rho}_{dm}$ distribution in the Solar system, then the
most stringent constraint ${\rho}_{dm} < 1.1\cdot10^{-20}$ g cm$^{-3}$
is obtained from the data for Saturn. The mass within
the spherical volume with the size of Saturn's orbit
is then $M_{dm} < 7.1\cdot10^{-11} M_{\odot}$, which is within the
error of the total mass of the main asteroid belt.

We can also consider the case where a continuous
gravitating medium has some concentration to the
Solar system center. Studies under the assumption of
density concentration to the center have already been
carried out, for example, in Fr\`{e}re et al. (2008). As a
model of the ${\rho}_{dm}$ distribution, we took the expression
\begin{equation}\label{f-6}
    {\rho}_{dm} = {\rho}_{0} \cdot e^{-cr} ,
\end{equation}
where ${\rho}_{0}$ is the central density and $c$ is a positive
parameter characterizing an exponential decrease in
density to the periphery. The value of $c=0$ corresponds
to a uniform density. Function (\ref{f-6}) is everywhere
finite, has no singularities at the center and
on the periphery, and is integrable. The expressions
for the gravitational potential for an inner point at
distance $r$ and the mass inside a sphere of radius $r$ 
for distribution (\ref{f-6}) are, respectively,
\begin{equation}\label{f-7}
    U(r) = 4\pi G {\rho}_{0} /r \cdot [2-e^{-cr}(cr+2)]/c^3  ,
\end{equation}
\begin{equation}\label{f-8}
 M_{dm} = 4\pi{\rho}_{0} [2/c^3 - e^{-cr}(r^2/c + 2r/c^2 + 2/c^3)] .
\end{equation}
In contrast to the potential $U(r)$ (\ref{f-7}), Eq. (\ref{f-8}) for the
mass $M_{dm}$ has no singularities for c --> 0, despite the
presence of $c^3$ in the denominator, and transforms into
the expression for a homogeneous sphere
  $$M(r)_{dm}= {4 \over 3}\pi r^3 {\rho}_{0}.$$

The values in Table~7 may be considered as estimates
of the dark matter density at various distances.
Indeed, if we take into account the fact that the dark
matter density is almost constant in a comparatively
narrow range of radial distances (the value of $c$ in (\ref{f-6})
is not too large), then the density of the extended
medium can be roughly assumed to be constant in the
range of $r$ due to the eccentricity of the planetary orbit.
Thus, when a changing density is considered, the estimates
from each planet may be considered as a local
estimate of ${\rho}_{dm}$ for the distance $r=a_{orb}$, where $a_{orb}$ is
the semimajor axis of the planetary orbit. Allowance
for the distributed dark matter $M_{dm}$ between the Sun
and the planetary orbit gives very small corrections
and contribution (in the tenths or elevenths decimal
digit) to the total attractive central mass determined
by the solar mass. Therefore, we can use Eq. (\ref{f-5a}) with
a sufficient accuracy and estimate the density ${\rho}_{dm}$
near the planetary orbit from the perihelion precession
produced by the dark matter.

When constructing Table~7 to estimate the dark
matter density, we took overestimated perihelion precessions
of planets corresponding to the errors of their
determination, i.e., the table contains constraining
upper limits. Using the data from it with similar
properties, we will obtain the density distribution (\ref{f-6}).
To find the parameters ${\rho}_{0}$ and $c$ in (\ref{f-6}), we took the
most reliable data in Table~7 for Saturn (${\rho}_{dm}<1.1\cdot 10^{-20}$ 
g cm$^{-3}$), Mars (${\rho}_{dm}<1.4\cdot10^{-20}$ g cm$^{-3}$),
and the Earth (${\rho}_{dm}<1.4\cdot10^{-19}$ g cm$^{-3}$). After the
arrival of the Cassini spacecraft to Saturn, a highly
accurate series of observations since 2004 has appeared.
For Mars, there is a large and long set of
observations related to spacecraft on its surface and
near it. Since the observations are performed from
the Earth, the improvement of the Earth's orbit is
based on all observations and includes measurements
of different accuracies.

For the Saturn-Mars pair, we obtained ${\rho}_{0}=1.50\cdot10^{-20}$ 
g cm$^{-3}$ and $ c=0.0279$~AU$^{-1}$. This
corresponds to a very flat density curve (\ref{f-6}). The
dark matter mass within the spherical volume corresponding
to Saturn's orbit turned out to be $M_{dm} < 7.6\cdot10^{-11} M_{\odot}$.

The Saturn-Earth pair gives ${\rho}_{0}=1.86\cdot10^{-19}$ g cm$^{-3}$ and 
$c=0.290$~AU$^{-1}$ and a steeper rise in ${\rho}_{dm}$ to the center, 
majorizing the density estimate for Mars. For these parameters, the mass
$M_{dm}$ within Saturn's orbit is $M_{dm} < 1.7\cdot10^{-10}  M_{\odot}$,
which is also within the error in the total mass of the
main asteroid belt ($\pm 2.1\cdot10^{-10}  M_{\odot}$).

The situation and the results did not change
greatly compared to the hypothesis of a uniform
density ${\rho}_{dm}$ -- the estimated total dark matter mass
within Saturn's orbit increased by a factor of $\sim 2.5$,
although the density distribution in the latter case
gives a significant increase to the center. Note that
a change in the parameter ${\rho}_{0}$ in the exponential
distribution (\ref{f-6}), just as in the density $\rho$ for a uniform
distribution, leads to the corresponding almost linear
change in the secular perihelion drift in the entire
range of distances from Mercury to Saturn. An
increase in the parameter $c$ causes the perihelion precession
to decrease in accordance with the decrease
in density ${\rho}_{dm}$ with distance $r$.

\bigskip
\centerline{CONCLUSIONS}
\smallskip

We investigated and estimated the possible gravitational
influence of dark matter in the Solar system
on the motion of planets based on the EPM2011
planetary ephemerides using about 677 thousand positional
observations of planets and spacecraft, most
of which belong to present-day ranging. Our results
show that the mass of the dark matter, if present,
and its density ${\rho}_{dm}$ are much lower than the presentday
errors in these parameters. We found that the
density ${\rho}_{dm}$ is less than $1.1\cdot10^{-20}$, $1.4\cdot10^{-20}$, and
$1.4\cdot10^{-19}$ g cm$^{-3}$ at the orbital distances of Saturn,
Mars, and the Earth, respectively. Taking into
account our constraining estimates, we considered
the case of a possible concentration of dark matter
to the Solar system center. The dark matter mass in
the sphere within Saturn's orbit should be less than
$1.7\cdot10^{-10} M_{\odot}$ even if its possible concentration is
taken into account.

\bigskip
\centerline{REFERENCES}
%\smallskip
 1. J. D. Anderson, E. L. Lau, A. H. Taylor, et al., Astrophys.J. {\bf342}, 539 
(1989).

 2. J. D. Anderson, E. L. Lau, T. P. Krisher, et al., Astrophys.J. {\bf448}, 885 
(1995).

 3. G. Bertone, D. Hooper, and J. Silk, Phys. Rep. {\bf405}, 279 (2005).

 4. A. Fienga, J. Laskar, P. Kuchynka, et al., Cel. Mech. Dyn. Astr. {\bf111}, 
363 (2011).

 5. C. Flynn, J. Sommer-Larsen, and P. R. Christensen, Mon. Not. R. Astron. Soc. 
{\bf281}, 1027 (1996).

 6. J.-M. Fr\`{e}re, F.-S. Ling, and G.Vertongen, Phys. Rev. D {\bf77}, 083005 
(2008).

 7. A. M. Fridman and A. V. Khoperskov, {\it Physics of Galactic Disks} (Fizmatlit, 
Moscow, 2011) [in Russian].

 8. M. W. Goodman and E. Witten, Phys. Rev. D {\bf31}, 3059 (1985).

 9. O. Gron and H. H. Soleng, Astrophys. J. {\bf456}, 445 (1996).

10. A. Hundhausen, {\it Coronal Expansion and Solar Wind} (Springer, Berlin, 
Heidelberg, New York, 1972; Mir, Moscow, 1976).

11. L. Iorio, J. Cosmol. Astropart. Phys. {\bf 05}, 002 (2006).

12. L. Iorio, J. Cosmol. Astropart. Phys. {\bf 05}, 018 (2010).

13. D. L. Jones, E. Fomalont, V. Dhawan, et al., Astron. J. {\bf141}, 29 (2011).

14. I. D. Karachentsev, Astrofizika {\bf2}, 81 (1966).

15. I. D. Karachentsev, Phys. Usp. {\bf44}, 818 (2001).

16. N. S. Kardashev, A. V. Tutukov, and A. V. Fedorova, Astron. Zh. {\bf82}, 157 
(2005).

17. R. Keisler, C. L. Reichardt, K. A. Aird, et al., Astrophys. J. {\bf743}, 28 
(2011).

18. I. B. Khriplovich, Int. J. Mod. Phys. D {\bf16}, 1475 (2007).

19. I. B. Khriplovich and E. V. Pitjeva, Int. J. Mod. Phys. D {\bf15}, 615 (2006).

20. E. Komatsu, K. M. Smith, J. Dunkleg, et al., Astrophys. J. Suppl. Ser. 
{\bf192}, 18 (2011).

21. A. S. Konopliv, S. W. Asmar, W. M. Folkner, et al., Icarus {\bf211}, 401 
(2011).

22. M. Kowalski, D. Rubin, G. Aldering, et al., Astrophys. J. {\bf686}, 749 (2008).

23. G. A. Krasinsky and M. V. Vasilyev, {/it Dynamics and Astrometry of Natural 
and Artificial Celestial Bodies, Proceedings of the IAU Coll. No. 165}, Ed. by 
I. M. Wytrzyszczak, J. H. Lieske, and R. A. Feldman (Kluwer Acad. Publ., 
Dordrecht, 1997), p. 239.

24. L. D. Landau and E. M. Lifshitz, {\it Course of Theoretical Physics, Vol. 1: 
Mechanics} (Pergamon Press, New York, 1988; Fizmatlit,Moscow, 2004).

25. J. Lundberg and J. Edsj\"{o}, Phys. Rev. D {\bf69}, 123505 (2004).

26. B. Luzum, N. Capitaine, A. Fienga, et al., Celest. \& Mech. Dyn. Astr. 
{\bf110}, 293 (2011).

27. K. L. Nordtwedt, Astrophys. J. {\bf437}, 529 (1994).

28. K. L. Nordtwedt, J. Mueller, and M. Soffel, Astron. Astrophys. {\bf293L}, L73 
(1995).

29. E. N. Parker, {\it Interplanetary Dynamical Processes} (Interscience, New York, 
1963; Mir, Moscow, 1965).

30. E. N. Parker, {\it Dynamical Theory of the Solar Wind}, Space Sci. Rev. {\bf4}, 
666 (1965).

31. A. Peter, Phys. Rev. D {\bf79}, 103531 (2009).

32. A. Peter, arXiv: astro-ph/1201.3942 (2012).

33. E. V. Pitjeva, Solar System Res. {\bf39}, 176 (2005).

34. E. V. Pitjeva, in {\it A Giant Step: From Milli- to Micro-Arcsecond Astrometry, 
Proceedings of the IAU Symposium No. 248}, Ed. byW. J. Jin, I. Platais, and 
M. A. C. Perryman (Cambridge Univ. Press, 2008), p. 20.

35. E. V. Pitjeva, in {\it Relativity in Fundamental Aastronomy, Proceedings of the 
IAU Symposium No. 261}, Ed. by S. Klioner, P. K. Seidelmann, and M. Soffel 
(Cambridge Univ. Press, 2010), p. 170.

36. E. V. Pitjeva and E. M. Standish, Celest. Mech. Dyn. Astron. {\bf103}, 365 (2009)

37. E. V. Pitjeva and N. P. Pitjev, Solar System Res. {\bf46}, 78 (2012).

38. M. Sereno and Ph. Jetzer,Mon. Not. R. Astron. Soc. {\bf371}, 626 (2006).

39. F. Zwicky, Helv. Phys. Acta {\bf6}, 110 (1933).

\end{document}